\documentclass[12pt,a4paper]{article}
\usepackage[utf8]{inputenc}
\usepackage[english]{babel}
\usepackage[left=2cm,right=2cm,top=2cm,bottom=2cm]{geometry}
\usepackage{array}
\newcolumntype{P}[1]{>{\centering\arraybackslash}p{#1}}
\usepackage{latexsym}
\usepackage{selinput}
\usepackage{latexsym}
\usepackage{amsmath}
\usepackage{amsfonts}
\usepackage{amssymb}
\usepackage{array}
\newcolumntype{P}[1]{>{\centering\arraybackslash}p{#1}}
\usepackage{caption}
\usepackage[colorlinks=true,linkcolor=blue]{hyperref}
\usepackage{graphics}
\usepackage{graphicx}
\usepackage{cite}

\title{Classical Planck Spectrum for Relative Thermal Radiation, Classical Zero-Point Radiation, and Scale Parameter}
\author{J. Tapia$^{1}$,H. González$^{2}$, and R. Rubiano$^{3}$\\
  \small Universidad Surcolombiana, Physical of Program, Neiva,  A.A 385, Colombia.\\
\date{Received: date / Revised version: date}
}
\providecommand{\keywords}[1]
{
  \small	
  \textbf{\textit{Keywords---}} #1
}
\begin{document}
\maketitle

\abstract{In this work we obtain Planck's blackbody spectrum from the thermal scalar radiation contained in a resonant cavity of volume V in the context of classical mechanics, which provides the classical zero-point electromagnetic radiation in terms of a scale parameter that depends on geometric properties of the enclosure and electrical magnitudes. The scale parameter of the classical zero-point electromagnetic radiation is associated to experimental measurements of Casimir forces, but we show that theoretically its value for known radiant cavities that approach a blackbody has a numerical value of the order of Planck's constant.
\\


\maketitle
\section{Introduction}
\label{intro}
The contributions of Rayleigh-Jeans \cite{J. H. Jeans,J. W. Rayleigh} and Planck \cite{1M. Planck} to elucidate the nature of the radiation emitted by hot bodies was elaborated before the year 1900, a time in which the special theory of relativity had not been formulated \cite{W. Thomson}. The foregoing indicates that this classical radiation law, and its quantum version in terms of Planck's constant does not contain space-time geometry explicitly, nor the Special Theory of Relativity \cite{2M. Planck}.  
The Zero-point radiation is included in quantum field theory \cite{M. Born,R. L. Jaffe,M. Bartelmann,S. Weinberg,P. W. Higgs}; however, this zero-point energy also emerges in classic relativistic electrodynamics and is determinant for explaining Casimir forces, Van der Waals forces, specific heats, and other phenomena apparently belonging to quantum mechanics \cite{S. K. Lamoreaux,M. Ibison,B. I. Ivlev}. 
\\

 There are derivations of blackbody thermal radiation spectrum that lead to Planck's law, with the classical zero-point energy, and that do not take into account the assumptions about quantization of energy \cite{1T. H. Boyer,2T. H. Boyer,3T. H. Boyer,4T. H. Boyer}. The foregoing would indicate that radiation emitted by hot bodies can also arise in a classical context, with the incorporation of the Special Theory of Relativity in the formulation of classical electrodynamics and the classical relativistic field theory. In reference to Planck's constant part of the scientific community discards its use as a universal constant, it can appear in diverse contexts \cite{5T. H. Boyer,F. W. Hehl,Y. L. Bolotin,L. Eaves}, and in this classical approximation, a parameter of the order of Planck's constant emerges, In this article, we incorporate to the Planck radiation law derived from classical relativistic electrodynamics and with zero-point radiation a scale parameter associated with the geometrical and electromagnetic characteristics of the thermal radiation contained in a finite enclosure of volume V. This scale parameter is associated with experimental measures of Casimir forces, in such a way that the structure of Planck's law differs from that generally presented in literature in terms of Planck's constant $\hbar $. 
\\

The organization of this work is distributed as follows: In the first section we present a synthesis of Rayleigh-Jeans and Planck radiation spectra. The second section contains the thermodynamics of normal radiation modes and the derivation of the Planck spectrum using classical physics, including classical zero-point radiation In the third section, which is the central part of our work, we present a version of the Planck spectrum for thermal radiation involving classical zero-point radiation, and the inclusion of scale parameter of zero-point energy. In the fourth section, we illustrate some specific examples of physical systems we apply the obtained Planck's radiation law to and determine the scale parameter value for each of them. In the last section we present an analysis of our results and their respective conclusions.

\section{Blackbody Radiaton}
\label{sec:1}

A blackbody absorbs and emits radiation perfectly, that is, there is no dominant frequency range. Therefore, the intensity of the radiation emitted is related to the amount of energy in the body in thermal equilibrium. The blackbody radiation theory development history is very interesting because it gave rise to the discovery of quantum theory \cite{J. Baggott}. The first experimental studies established that the emissivity of a blackbody is a function of frequency and temperature. A measure of emissivity may be the quantity $\rho (v,T)$ which is the density of radiated energy per volume unit and per frequency unit at an absolute temperature T and at a frequency $v$. The first theoretical studies used Maxwell's equations for calculating the electromagnetic modes and for determining $\rho (v,T)$ density of modes. For example, Wilhelm Wien in 1896 used a simple model to derive the expression:

\begin{equation}
\rho (v,T)=av^{ 3 }\operatorname{exp}\left[ -bv^{ 3 } \right] 
\end{equation}

Where $a,b$  were constant. However, the above equation fails in the range of experimental data low frequencies. In 1900 Lord Rayleigh published a model based on the modes of electromagnetic waves of a cavity. Each mode had a particular frequency and can give or receive energy in a continuous way. The total number of modes per volume, $N(v)$, is \cite{L. D. Landau}: 

\begin{equation}
\frac { dN(v) }{ dV } =\frac { 8\pi v^{ 2 } }{ c^{ 3 } } 
\end{equation}

Rayleigh assigned an energy ${ { k }_{ B } }/{ 2 }$  to each electromagnetic mode, and ${ { k }_{ B } }/{ 2 }$  for the oscillation of the electric field, where ${ k }_{ B }=1.38066\quad { K }/{ J }$ is Boltzmann constant. In this way, the density of electromagnetic energy by frequency $\rho (v,T)$ becomes:

\begin{equation}
\rho (v,T)=\frac { dN(v) }{ dV } { k }_{ B }T=\frac { 8\pi v^{ 2 } }{ c^{ 3 } } { k }_{ B }T
\end{equation}

The last equation is known as the Rayleigh-Jeans distribution of blackbody radiation and fails dramatically in the ultraviolet part of the spectrum (historically referred to as the "ultraviolet catastrophe") \cite{R. Resnick}. It has been suggested that Planck discovered his radiation formula on the afternoon of October 7th, 1900. Planck had taken into account some additional experimental data from Heinrich Reubens and Ferdinand Kurlbaum, as well as Wien's formula, and he deduced an expression that fit all the experimental data available. His formula was the one now known as the blackbody radiation formula given by:

\begin{equation}
\rho (v,T)=\frac { 8\pi v^{ 2 } }{ c^{ 3 } } \frac { hv }{ { \operatorname{exp}\left[ \frac { hv }{ { k }_{ B }T }  \right] -1 } }   
\end{equation}

Where $h=6,626{ \times  }10^{ -34 }$  Joule.seconds is known as Planck's constant. The above expression is reduced to the Wien formula for high frequencies (i.e.,${ hv }/{ k_{ B }T }\gg 1$) and Rayleigh-Jeans formula for low frequencies (i.e., ${ hv }/{ k_{ B }T }\ll 1$).

\section{Planck Spectrum with Classic Physics and zero-point energy}
\label{sec:2}

A small harmonic oscillator reaches equilibrium with thermal radiation at the same average energy as a normal mode of radiation at the same frequency as the oscillator \cite{V. H. Lavenda}. Alternatively, we can think that a radiation mode behaves like a harmonic oscillator. 
\\

The normal mode of radiation thermodynamics has two thermodynamic variables T and $\omega$, like a harmonic oscillator, and takes a particularly simple form \cite{V. H. Lavenda}. The average energy for the normal mode is related to the canonical potential function $\phi (\omega /T)$  by means of the equation:

\begin{equation}
U(\omega ,T)=-\omega \phi '(\omega /T)
\end{equation}

The first law of thermodynamics is:

\begin{equation}
dQ=dU+dW
\end{equation}
\\

The change of dS entropy is related to temperature T by:

\begin{equation}
dS=\frac { dQ }{ T } 
\end{equation}
\\

From the laws of thermodynamics, it is inferred that the energy for each normal mode is expressed by:

\begin{equation}
U=\omega f\left( \frac { \omega  }{ T }  \right) 
\end{equation}
\\

$f\left( \frac { \omega  }{ T }  \right)$ is an unknown function. The above equation is known as Wien's theorem. \cite{1T. H. Boyer} Helmholtz free energy is related to $U(\omega ,T)$  energy by:
\\

\begin{equation}
F\left( \omega ,T \right) =U\left( \omega ,T \right) -TS
\end{equation}
\\

\begin{equation}
U\left( \omega ,T \right) =F\left( \omega ,T \right) +TS
\end{equation}
\\

This random radiation that exists at temperature $T=0$ is a classical zero-point electromagnetic radiation. Thus, to account for Casimir forces observed experimentally, and for each normal mode, the average energy becomes:
\\

\begin{equation}
U\left( \omega ,0 \right) =F\left( \omega ,0 \right) =\frac { \hbar \omega  }{ 2 }   
\end{equation}
\\

Derivations of Planck's blackbody spectrum that occur within classical mechanics, include classical electromagnetic zero-point radiation and thermodynamic ideas. Thermodynamics includes the first and second laws, thermodynamic potential, Helmholtz free energy, and Wien's theorem. In the mathematical context, the classic scalar theory of relativistic field, the accelerated frames of Rindler, and the smooth monophonic behavior of thermodynamic functions are used. \cite{S. K. Lamoreaux,M. Ibison,B. I. Ivlev,1T. H. Boyer,2T. H. Boyer,3T. H. Boyer,4T. H. Boyer}; Planck's blackbody spectrum is:
\\

\begin{equation}
U(\omega ,T)=\frac { \hbar \omega  }{ \operatorname{exp}\left[ \frac { \hbar \omega  }{ { k }_{ B }T }  \right] -1 } +\frac { \hbar \omega  }{ 2 }   
\end{equation}
\\

\section{Planck spectrum for relativistic classical scalar radiation, classical zero-point radiation, and scale parameter.}
\label{sec:3}

Consider the electromagnetic radiation contained in an enclosure of volume V in the free space. Assuming that the contributions of electric and magnetic parts to energy density are equal, in these circumstances in SI units:

\begin{equation}
u=\frac { 1 }{ 2 } { \varepsilon  }_{ 0 }E^{ 2 }+\frac { 1 }{ 2{ \mu  }_{ 0 } } B^{ 2 }={ \varepsilon  }_{ 0 }E^{ 2 }
\end{equation}
\\

With the equation of motion for electromagnetic field given by:

\begin{equation}
\frac { 1 }{ { c }^{ 2 } } \frac { { \partial  }^{ 2 }\phi  }{ \partial { t }^{ 2 } } -\frac { { \partial  }^{ 2 }\phi  }{ \partial { x }^{ 2 } } -\frac { { \partial  }^{ 2 }\phi  }{ \partial { y }^{ 2 } } -\frac { { \partial  }^{ 2 }\phi  }{ \partial { z }^{ 2 } } =0
\end{equation}
\\

Where Lagrangian density is inferred from:

\begin{equation}
\L ={ \varepsilon  }_{ 0 }\left[ \frac { 1 }{ { c }^{ 2 } } { \left( \frac { { \partial  }\phi  }{ \partial { t } }  \right)  }^{ 2 }-{ \left( \frac { { \partial  }\phi  }{ \partial { x } }  \right)  }^{ 2 }-{ \left( \frac { { \partial  }\phi  }{ \partial { y } }  \right)  }^{ 2 }-{ \left( \frac { { \partial  }\phi  }{ \partial { z } }  \right)  }^{ 2 } \right] 
\end{equation}
\\

And consequently the energy:

\begin{equation}
U={ \varepsilon  }_{ 0 }\int { \left[ \frac { 1 }{ { c }^{ 2 } } { \left( \frac { { \partial  }\phi  }{ \partial { t } }  \right)  }^{ 2 }-{ \left( \frac { { \partial  }\phi  }{ \partial { x } }  \right)  }^{ 2 }-{ \left( \frac { { \partial  }\phi  }{ \partial { y } }  \right)  }^{ 2 }-{ \left( \frac { { \partial  }\phi  }{ \partial { z } }  \right)  }^{ 2 } \right] dV } 
\end{equation}
\\

Random thermal movements can be expressed in terms of the oscillations of the normal modes with random phases. The field expressed in electromagnetic modes is given by the expansion term \cite{A. Zanglwill}.
\\

\begin{equation}
\phi \left( ct,x,y,z \right) =\sum _{ m=-\infty  }^{ \infty  }{  } \sum _{ n=-\infty  }^{ \infty  }{  } \sum _{ l=-\infty  }^{ \infty  }{ \frac { g\left( c\vec { k }  \right)  }{ { \left( V \right)  }^{ { 1 }/{ 2 } } } \cos { \left[ \vec { k } .\vec { r } -kct-\theta \left( \vec { k }  \right)  \right]  }  }   
\end{equation}
\\

Where $g\left( c\vec { k }  \right) $  is called amplitude or distribution function by mode. Thus, the energy of thermal radiation in a cavity can be expressed as a sum over the energies of the normal modes of oscillation. Within classical physics, thermal radiation is treated as classical electromagnetic radiation defining:
\\

\begin{equation}
g\left( c\vec { k }  \right) ={ \delta  }_{ 0 }h\left( c\vec { k }  \right) 
\end{equation}
\\

With ${ \delta  }_{ 0 }$  as a constant factor extracted from $g\left( c\vec { k }  \right) $, using eqs. (17) and (18) in eq. (16), electromagnetic energy U is Each oscillation mode will have an energy given by:
\\

\begin{equation}
U={ \varepsilon  }_{ 0 }{ { \delta  }_{ 0 } }^{ 2 }\sum _{ m=-\infty  }^{ \infty  }{  } \sum _{ n=-\infty  }^{ \infty  }{  } \sum _{ l=-\infty  }^{ \infty  }{ { \left[ h\left( c\vec { k }  \right)  \right]  }^{ 2 }{ k }^{ 2 }\cos { \left[ \vec { k } .\vec { r } -kct-\theta \left( \vec { k }  \right)  \right]  }  } 
\end{equation}
\\

Each oscillation mode will have an energy given by:
\\

\begin{equation}
{ U }_{ k }={ \varepsilon  }_{ 0 }{ { \delta  }_{ 0 } }^{ 2 }{ k }^{ 2 }{ \left[ h\left( c\vec { k }  \right)  \right]  }^{ 2 }
\end{equation}
\\

For eq. (18) to be dimensionally correct, and given that the term ${ k }^{ 2 }{ \left[ h\left( c\vec { k }  \right)  \right]  }^{ 2 }$  has units of volume, ${ \delta  }_{ 0 }$  must have electric field dimensions; 
\\

\begin{equation}
{ U }_{ k }={ \varepsilon  }_{ 0 }{ E_{ 0 } }^{ 2 }{ k }^{ 2 }{ \left[ h\left( c\vec { k }  \right)  \right]  }^{ 2 }
\end{equation}
\\

Thus In eq. (22) has been taken ${ { \delta  }_{ 0 } }^{ 2 }={ E_{ 0 } }^{ 2 }$. Being ${ E_{ 0 } }^{ 2 }$   the electric field average square in the proximity of the radiant enclosure. It is possible to adapt the units in the ${ k }^{ 2 }{ \left[ h\left( c\vec { k }  \right)  \right]  }^{ 2 }$  coefficient by extracting a constant factor having units of volume per time. Assuming that the constant factor is:
\\

\begin{equation}
{ \alpha  }_{ 0 }=\frac { { L }^{ 4 } }{ c } 
\end{equation}
\\

Where L is a factor with length dimensions and c is the speed of light in free space:
\\

\begin{equation}
{ \left[ h\left( c\vec { k }  \right)  \right]  }^{ 2 }={ \alpha  }_{ 0 }{ \left[ f\left( c\vec { k }  \right)  \right]  }^{ 2 }
\end{equation}
\\

Thus, energy by mode acquires the form:
\\

\begin{equation}
{ U }_{ k }={ \varepsilon  }_{ 0 }{ E_{ 0 } }^{ 2 }\frac { { L }^{ 4 } }{ c } { k }^{ 2 }{ \left[ f\left( c\vec { k }  \right)  \right]  }^{ 2 }=\beta { k }^{ 2 }{ \left[ f\left( c\vec { k }  \right)  \right]  }^{ 2 }={ k }^{ 2 }{ \left[ F\left( c\vec { k }  \right)  \right]  }^{ 2 }
\end{equation}
\\

The parameter $\beta $ has been inserted in the function ${ \left[ F\left( c\vec { k }  \right)  \right]  }^{ 2 }$, so that:
\\

\begin{equation}
{ \left[ F\left( c\vec { k }  \right)  \right]  }^{ 2 }=\beta { \left[ f\left( c\vec { k }  \right)  \right]  }^{ 2 }
\end{equation}
\\

We have found that parameter $\beta$ has units of J.S, and is defined by:
\\

\begin{equation}
\beta ={ \varepsilon  }_{ 0 }{ E_{ 0 } }^{ 2 }\frac { { L }^{ 4 } }{ c } 
\end{equation}
\\

$\beta$ can be determined numerically by knowing the linear dimension L of the enclosure and the value of ${ E_{ 0 } }^{ 2 }$  which is the average value of the electric field amplitude, which corresponds to different modes of oscillation. The parameter $\beta$ fixed, in this approximation, the classical zero-point energy.
\\

\begin{equation}
{ U }_{ 0 }=\frac { 1 }{ 2 } \beta \omega 
\end{equation}
\\

From ${ U }_{ k }$  in eq. (24), energy per oscillation mode, we can determine the values of the criterion of smooth monotonic behavior of the canonical partition function $\phi \left( { \omega  }/{ T } \right)$  at high and low temperature, to find Planck's radiation spectrum with classic zero-point energy \cite{4T. H. Boyer}. For small T there is the classic zero-point energy ${ 1 }/{ 2 }\beta \omega$ , from eq. (5); 
\\

\begin{equation}
{ U }_{ 0 }=-\omega { \phi  }_{ 0 }'\left( \frac { \omega  }{ T }  \right) =\frac { 1 }{ 2 } \beta \omega 
\end{equation}
\\

Taking into account that ${ \omega  }/{ T }=z$:
\\

\begin{equation}
-{ \phi  }_{ 0 }'\left( z \right) \rightarrow \frac { \beta  }{ 2 } \quad ;\quad \quad z\rightarrow \infty 
\end{equation}
\\

As follows:
\\

\begin{equation}
-{ \phi  }_{ 0 }\left( z \right) \rightarrow \frac { \beta  }{ 2 } z\quad ;\quad \quad z\rightarrow \infty 
\end{equation}
\\

For large values of T, Rayleigh-Jeans limit, eq. (5) gives:
\\

\begin{equation}
{ U }_{ \infty  }=-\omega { \phi  }_{ \infty  }'\left( z \right) ={ k }_{ B }T
\end{equation}
\\

When,

\begin{equation}
-{ \phi  }_{ \infty  }'\left( z \right) \rightarrow \frac { { k }_{ B } }{ z } \quad ;\quad \quad z\rightarrow 0
\end{equation}
\\

\begin{equation}
-{ \phi  }_{ \infty  }\left( z \right) \rightarrow { k }_{ B }\ln { z } \quad ;\quad \quad z\rightarrow 0
\end{equation}
\\

Where ${ k }_{ B }$  is Boltzmann constant Eqs. (30) and (33) give extreme behaviors of the canonical potential function ${ \phi  }\left( z \right) $. In general, the canonical potential function ${ \phi  }\left( z \right) $, like its derivatives, must preserve its characteristic smooth monotonicity throughout the range $\left( 0,\infty  \right)$; the quotient:

\begin{equation}
y=\frac { { U }_{ 0 } }{ { U }_{ \infty  } } =\frac { \frac { 1 }{ 2 } \beta \omega  }{ { k }_{ B }T } =\frac { \beta \omega  }{ { 2k }_{ B }T } =\frac { \beta  }{ { 2k }_{ B } } z
\end{equation}
\\

Is the argument of the smooth monotonic function $M\left( y \right)$ , with extreme values given by eqs (29) and (32), which determine energy $U(\omega ,T)$.

\begin{equation}
U(\omega ,T)=-\omega { \phi  }\left( \frac { \omega  }{ T }  \right) =-\omega M\left( y \right) 
\end{equation}
\\

In reference \cite{4T. H. Boyer} the author demonstrated that an appropriate function that maintains the smooth monotony of ${ \phi  }\left( z \right)$ and its derivatives, the function of interpolation between these two limits, in the entire range of variation of ${ \omega  }/{ T }=z\quad $  is given by the equation:
\\

\begin{equation}
{ \phi  }\left( z \right) =-\beta \ln { \left( \sinh { z }  \right)  } 
\end{equation}
\\

Thus,

\begin{equation}
{ \phi  }'\left( \frac { \omega  }{ T }  \right) =M\left( y \right) =-\frac { \beta  }{ 2 } \coth { \left( \frac { \beta \omega  }{ 2{ k }_{ B }T }  \right)  } 
\end{equation}
\\

Adapting this function to eq. (35):

\begin{equation}
U(\omega ,T)={ U }_{ 0 }\coth { \left( \frac { \beta \omega  }{ 2{ k }_{ B }T }  \right)  } =\frac { 1 }{ 2 } \beta \omega \coth { \left( \frac { \beta \omega  }{ 2{ k }_{ B }T }  \right)  } 
\end{equation}
\\

\begin{equation}
U(\omega ,T)=\frac { \beta \omega  }{ \operatorname{exp}\left[ \frac { \beta \omega  }{ { k }_{ B }T }  \right] -1 } +\frac { \beta \omega  }{ 2 }   
\end{equation}
\\

Eq. (38) is a version of Planck's blackbody spectrum, which includes the classical zero-point energy ${ U }_{ 0 }$, and scale parameter $\beta$ . The classic zero-point energy is given by:
\\

\begin{equation}
{ U }_{ 0 }=\frac { 1 }{ 2 } \beta \omega 
\end{equation}
\\

Where the scale parameter is:

\begin{equation}
\beta =\frac { { \varepsilon  }_{ 0 } }{ c } { \left[ { E_{ 0 } }{ L }^{ 2 } \right]  }^{ 2 }
\end{equation}
\\

The scale parameter depends on geometric properties, linear dimension L of the enclosure, and electromagnetic properties, permittivity of the free space ${ \varepsilon  }_{ 0 }$, speed of light c, and the square of the amplitude of the electric field average value.

\section{Scale parameter and its connection with radiant physical systems.}
\label{sec:4}

Scale parameter sets zero-point classical electromagnetic energy and has J.S dimensions in the international system of units:
\\

\begin{equation}
\beta =\frac { { \varepsilon  }_{ 0 } }{ c } { \left[ { E_{ 0 } }{ L }^{ 2 } \right]  }^{ 2 }\cong 2.95{ \times 10 }^{ -20 }\frac { { C }^{ 2 }s }{ N{ m }^{ 3 } } \left( { \left[ { E_{ 0 } }{ L }^{ 2 } \right]  }^{ 2 } \right) 
\end{equation}
\\

With the linear dimension L in meters and $E_{ 0 }$  in ${ V }/{ m }$ or in ${ N }/{ C }$. Taking into account that the parameter $\beta $ is related to experimental measurements of Casimir forces and that its order of magnitude is similar to $\hbar$ , an estimation of the linear dimension of some radiant systems reported in the literature \cite{1C. Schiller} is made to observe if they are consistent with their real values. Table \ref{tab:1} shows some of the values reported for $E_{ 0 }$  and the estimate made for the linear dimension L using eq. (41) where ${ \beta  }/{ 2 }=0.527{ \times 10 }^{ -34 }J.s$ is the experimental value reported by Casimir Effect measurements \cite{2C. Schiller} in the table \ref{tab:2}.
\\
 
Using eq. (41) and the experimental value of ${ \beta  }/{ 2 }$  we can determine the typical values of linear dimension L for radiation systems. In Table \ref{tab:1}, in the third column we include the estimated values of L, In Table \ref{tab:2} in column 2  the experimental value of L, and in the third column the size of the equivalent physical system.
\\

\begin{table}
\centering
\caption{theoretical values of the linear dimension L and electric field $E_{ 0 }$.}
\label{tab:1}       
\begin{tabular}{P{5cm} P{5cm} P{5cm}}

\hline\noalign{\smallskip}

Radiant System & Values of  $E_{ 0 }\left( \frac { N }{ C }  \right)$ (a) & Theoretical Value $L (m)(b)$ \\
\noalign{\smallskip}\hline\noalign{\smallskip}
Field just beneath a high power line & 0.5 & $0.329{ \times 10 }^{ -3 }$ \\
Field of a 100 W bulb at 1 m distance & 50 & $34.570{ \times 10 }^{ -6 }$ \\
Electric fields in laser pulses & ${ 10 }^{ 13 }$ & $77.312{ \times 10 }^{ -12 }$ \\
Electric fields in  ${ U }^{ 91 }$  ions & ${ 10 }^{ 18 }$ & $0.244{ \times 10 }^{ -12 }$  \\
Maximum practical electric field in vacuum, limited by electron  pair production & $1.3\times { 10 }^{ 18 }$ & $0.214{ \times 10 }^{ -12 }$ \\
Maximum electric field in nature & $1.9\times { 10 }^{ 62 }$ & $1.744{ \times 10 }^{ -35 }$ \\

\noalign{\smallskip}\hline
\end{tabular}
 \caption*{{\small \textit{Note: $(a)$Data taken from the text:  C. Schiller , Motion Mountain The adventure of Physics Volume I. Light, Charges and Brains (Creative Commons, Munich, Germany, 2011).}}}
 \caption*{{\small \textit{Note: $(b)$Data taken from the text:  C. Schiller , Motion Mountain; The adventure of Physics Volume III Light, Charges and Brains, 30th ed. (Creative Commons, Munich, Germany, 2018).}}}
\label{Note:1}
\end{table}

\begin{table}
\centering
\caption{Determination of the linear dimension L average values from equation (41) which sets the experimental value of the scale parameter ${ \beta  }$.}
\label{tab:2}       
\begin{tabular}{P{5cm} P{5cm} P{5cm}}

\hline\noalign{\smallskip}

Radiant System & Experimental Value L (m) & Correspondence according to the value of L  (c) \\
\noalign{\smallskip}\hline\noalign{\smallskip}
Field just beneath a high power line & $0.300{ \times 10 }^{ -3 }$ & Unicellular protozoon of the amoeba genus \\
Field of a 100 W bulb at 1 m distance & $30{ \times 10 }^{ -6 }$ & Diameter of a human hair \\
Electric fields in laser pulses & $79{ \times 10 }^{ -12 }$ & Sodium atomic radio \\
Electric fields in  ${ U }^{ 91 }$  ions & $2.426{ \times 10 }^{ -12 }$ & Atomic nuclei size  \\
Maximum practical electric field in vacuum, limited by electron  pair production & $2.426{ \times 10 }^{ -12 }$ & Compton wavelength in electrons \\
Maximum electric field in nature & $1.62\times { 10 }^{ -35 }$ & Planck length \\

\noalign{\smallskip}\hline
\end{tabular}
\caption*{{\small \textit{Note: $(c)$Data taken from the text:  C. Schiller , Motion Mountain The adventure of Physics Volume I. Light, Charges and Brains (Creative Commons, Munich, Germany, 2011).}}}
\label{Note:2}
\end{table}

\section{Analysis and conclusions.}
\label{sec:5}

We have found the blackbody radiation spectrum in connection with zero-point energy using only classical physics, given by eq. (38) and introducing a scale parameter, which is defined in eq. (40) consequently, we have shown that the theoretical numerical value depends on the geometrical properties of the resonant box and the electrical properties according to our approximation.

Therefore, the scale parameter obtained for zero-point radiation energy was found with arguments of the classical theory and does not imply the quantization of energy; however, its numerical value is of the order of the Planck constant, since the typical electric fields generated by various oscillating physical systems have values consistent with the theoretical estimation sizes of the radiant system (see column 2 of Table \ref{tab:1}).

In conclusion, the scale parameter found is related to the experimental measurements of Casimir forces; In addition, the above arguments show that the Planck constant can appear in quantum mechanics or in classical mechanics, just as the zero-point of energy can be classical or quantum.

\section*{Acknowledgement}
We thank Professor Ricardo Gaitán for reading the manuscript and making important suggestions. We appreciate the total support of Surcolombiana University through the research project 2334.

%
%
%

%
%

\end{document}